# Disentangling electronic transport and hysteresis at individual grain boundaries in hybrid perovskites via automated scanning probe microscopy

Yongtao Liu,[1,a] Jonghee Yang,[2] Benjamin J Lawrie,[1,3] Kyle P. Kelley,[1] Maxim Ziatdinov,[1,4] Sergei V. Kalinin,[2,b] and Mahshid Ahmadi[2,c]

[1] Center for Nanophase Materials Sciences, Oak Ridge National Laboratory, Oak Ridge, TN 37830, USA

[2] Department of Materials Science and Engineering, University of Tennessee, Knoxville, TN, 37996, USA

[3] Materials Science and Technology Division, Oak Ridge National Laboratory, Oak Ridge, TN 37830, USA

[4] Computational Sciences and Engineering Division, Oak Ridge National Laboratory, Oak Ridge, TN 37830, USA

Underlying the rapidly increasing photovoltaic efficiency and stability of metal halide perovskites (MHPs) is the advance in the understanding of the microstructure of polycrystalline MHP thin film. Over the past decade, intense efforts have aimed to understand the effect of microstructure on MHP properties, including chemical heterogeneity, strain disorder, phase impurity, etc. It has been found that grain and grain boundary (GB) are tightly related to lots of microscale and nanoscale behavior in MHP thin film. Atomic force microscopy (AFM) is widely used to observe grain and boundary structures in topography and subsequently to study the correlative surface potential and conductivity of these structures. For now, most AFM measurements have been performed in imaging mode to study the static behavior, in contrast, AFM spectroscopy mode allows us to investigate the dynamic behavior of materials, e.g. conductivity under sweeping voltage. However, a major limitation of AFM spectroscopy measurements is that it requests manual operation by human operators, as such only limited data can be obtained, hindering systematic investigations of these microstructures. In this work, we designed a workflow combining the conductive AFM measurement with a machine learning (ML) algorithm to systematically investigate grain boundaries in MHPs. The trained ML model can extract GBs locations from the topography image, and the workflow drives the AFM probe to each GB location to perform a current-voltage (IV) curve automatically. Then, we are able to IV curves at all GB locations, allowing us to systematically understand the property of GBs. Using this method, we discover that the GB junction points are more photoactive, while most previous works only focused on the difference between GB and grains.

---

[a] liuy3@ornl.gov
[b] sergei2@utk.edu
[c] mahmadi3@utk.edu

Metal halide perovskites (MHPs) have become the most promising class of materials for the next generation of photovoltaic technology due to their rapidly increasing photovoltaic efficiency and low-cost manufacturing. In about a decade of development, the photovoltaic power conversion efficiency (PCE) of single junction and tandem MHP solar cells has surpassed 25% and 31%, respectively, comparable to silicon solar cells that underwent a much longer time period of development.[1] From the perspective of commercializing MHP photovoltaics, further research efforts are dedicated to boosting photovoltaic efficiency and improving the stability of MHP solar cells.[2-4] Underlying the rapidly increasing photovoltaic efficiency and stability is the advance in understanding of the microstructure of polycrystalline MHP thin film.[5-7] As such, intense efforts have aimed to elucidate the effect of microstructure on MHP properties.[8-15]

Microscale and nanoscale chemical heterogeneity has proven to be critical to microscale MHP behavior. Using synchrotron X-ray fluorescence microscopy, it was revealed that halide homogenization (in contrast to halide segregation at the micrometer scale) upon the addition of cesium iodide in mixed halide and mixed cation MHPs results in spatially homogeneous photogenerated carrier dynamics and long charge carrier lifetime, along with improved photovoltaic efficiency.[16] Multimodal microscopy measurements revealed that nanoscale compositional disorder can lead to electronic disorder and charge carrier funneling, which ultimately dominates the local optoelectronic response.[7] In addition, it was shown that nanoscale phase impurities act as both traps for photogenerated charge carriers and sites of photochemical degradation.[6] Studies with secondary ion mass spectrometry unveiled the micrometer level chemical redistribution and degradation under light and electric field, which affects the surface potential and electrical current in the MHP films.[17-21]

Grain structure and local strain gradients have also proven to substantially affect MHP optoelectronic properties. Crystallographic twin domains in MHPs have been intensively investigated with electron microscopy,[22] scanning probe microscopy,[12, 23-27] and optical microscopy.[28, 29] The crystallographic difference between adjacent domains was revealed by both transmission electron microscopy and electron backscatter diffraction.[22, 30] There still is controversy over whether this twin domain structure is a ferroelectric domain or not and how it affects the optoelectronic behavior of the materials. Theoretical simulations have shown the effect of such domain structures based on both ferroelectric nature and ferroelastic nature. In investigations with scanning photocurrent microscopy, it was shown that such twin structures potentially do not affect the charge carrier transport.[28] In contrast, confocal photoluminescence (PL) microscopy indicates the PL intensity variation between adjacent domains, which was attributed to the difference in local charge carrier concentration.[28, 31] In addition, PL microscopies have shown that the domain walls can act as shallow energetic barriers to delay charge carrier diffusion.[13] A structure akin to the twin domain is the subgrain boundary, which provides nonradiative recombination sites and restricts carrier diffusion.[32]

Understanding the chemical heterogeneity and crystallographic behavior in perovskite grains and grain boundaries is critical to the development of MHP solar cells with high efficiency and stability. Photovoltaic efficiency is determined by photocarrier generation and transport, and grain boundaries can serve as energy barriers and recombination sites that influence charge carrier transport and recombination. The stability of MHP solar cells is determined by both intrinsic material properties and extrinsic effects.[33, 34] One intrinsic property that affects MHP stability is ion migration.[35, 36] Thus, grain boundaries acting as ion migration highways play crucial roles in MHP stability. Extrinsic effects such as moisture and oxygen can penetrate into MHP films through grain boundaries and ultimately decompose MHP films.

Two popular tools for investigating grain boundary are scanning electron microscopy (SEM) and atomic force microscopy (AFM). SEM can be used for preliminary observation of grain and grain boundary structure, and grain size and boundary density are usually linked to photovoltaic device performance. AFM, particularly functional AFM such as Kelvin Probe Force Microscopy (KPFM) and conductive AFM (cAFM), can be used to observe grain and boundary structures in topography and to perform correlative studies of surface potential and conductivity of these structures. For instance, research combining KPFM and cAFM revealed spatial heterogeneity in short-circuit current and open-circuit voltage correlated with different crystal facets within individual grains, implying a direct impact of grain facets on photovoltaic efficiency.[37] KPFM studies showed charge accumulation at grain boundaries.[38-40] Tomographic and conductive AFM has revealed grain boundaries as highly interconnected conducting channels for carrier transport.[41] However, most AFM measurements on MHP have been performed in imaging mode, and they have only provided information at a static condition. AFM spectroscopy mode allows us to investigate the dynamic behavior of materials, e.g. conductivity under sweeping voltage, aka current-voltage curves.

Until now, AFM spectroscopy measurements have been limited by manual operation by human operators. For instance, an AFM probe can be manually positioned at a grain or a grain boundary to perform current-voltage measurements.[42] As a result, only very limited data have been obtained, and systematic investigations of objects of interest (e.g., mapping current-voltage measurements across a complete grain boundary) has been virtually impossible. Recently, we developed a deep convolutional neural network (DCNN) with a combination of a holistically nested edge detector and a deep residue learning framework to convert raw microscopy images into the segmented image showing objects of interest[43] (e.g., convert topography image to grain boundary image). Implementing this DCNN in an operating AFM will allow us to convert stream data to a segmented objects-of-interest image.

Here, we designed a workflow combining the conductive AFM measurement with above described DCNN to systematically investigate grain boundaries in MHPs. We trained the DCNN with pre-acquired data to convert a topography image to a grain boundary (GB) image. Then, the coordinates of GBs can be extracted from the GB image, and the workflow drives the AFM probe to each GB coordinate and performs a current-voltage curve (IV) in an automated manner without human intervention. At the end, the results including IV curves at all GB locations allow us to systematically understand the property of GBs.

As a model system, a state-of-the-art mixed formamidinium (FA) and cesium (Cs) cation (FACs) MHP (FACs-MHP), with 17% Cs composition ratio, is selected for this measurement. Mixed FA and Cs MHPs show excellent phase stability compared to the MHPs with volatile methylammonium (the MHP structures are compromised upon methylammonium evaporation) and a PCE of 22.7% has been reported for FACs-MHPs.[44, 45]

To assess the spatial heterogeneity of the MHP film, we performed hyperspectral cathodoluminescence (CL) microscopy first. Shown in Figure 1a is a scanning electron microscopy (SEM) image obtained during the CL measurement, in which grain and GB structures are visible. The local CL spectra, with a 60-nm pixel size, are simultaneously collected, and the hyperspectral CL image illustrates the spatial optoelectronic properties of this region. We performed non-negative matrix factorization (NMF) to decompose the properties encoded in the CL spectra, revealing local electronic features. Using five components, key spectral features and associated CL maps are successfully disentangled (as shown in Figure S1). We find that component#1 and component#2 likely correlate to CL emission at grains and GBs including junction points, respectively. These decomposed CL maps are shown in Figure 1b and 1c, and the corresponding CL spectra of both components, as well as the global-averaged CL spectrum are shown in Figure 1d. The CL spectrum of component#1 exhibits a main peak centered at 783 nm, consistent with the global averaged spectrum. The CL spectrum of component#2 includes a main peak around 756 nm (18 nm blue shifted from the global average), indicating slightly larger bandgap at these regions compared to that of grains. The spectrum also contains another small peak at around 505 nm, suggesting component#2 which is attributed to decomposed $PbI_2$ luminescence. Moreover, it is obvious that the intensity of the component#2 is higher at GBs and possibly even higher at some junction points, as shown in Figure 1e.

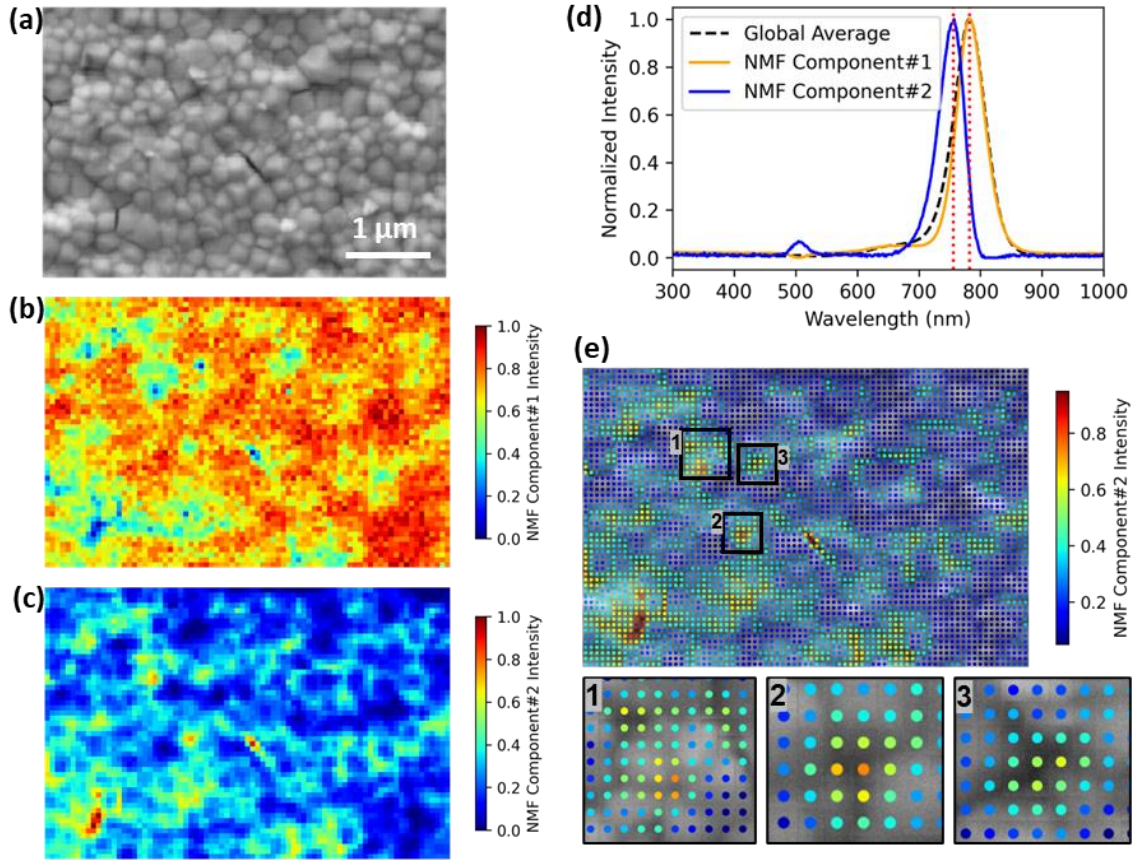

Figure 1. CL results. (a) SEM image shown the grain and GBs structures. (b-c) images of NMF component#1 and #2, respectively. (d) NMF component#1 and #2 spectral compared to the global average spectrum. (e) NMF component#2 distribution map over SEM image, where we showed three enlarged images indicating the higher component#2 intensity at GB junction points.

The above results lead us to investigating the electronic properties of junction points. Various microscopes are known as powerful tools for nanoscale characterization, for instance, we can conduct cAFM to study the nanoscale conductivity of functional materials. cAFM can be run in image mode and spectroscopy mode. In the image mode, we apply a DC voltage through the AFM tip and probe the current at each pixel. This allows us to obtain a current map showing the conductivity of the measured regions. In spectroscopy model, we can move the probe to the location of interest and apply a DC waveform to probe the current-voltage (IV) curve, the IV curve can offer more details about the sample. However, the limitations of these standard microscopic measurements are: image mode only allows to find the conductivity under a static condition, i.e, under a specific DC voltage; spectroscopy mode requires manual operation to select the measurement pixels (as shown in Figure 2a), which is monotonous and time-consuming. Traditionally, we are also able to perform grid spectroscopy measurement (e.g., shown in Figure 1 of CL grid spectroscopy measurements). However, in such grid spectroscopy measurements, the measurement points do not necessarily fall on the object of interests. In addition, a dense grid spectroscopy takes too long and potentially results in materials damage or degradation. Therefore, it necessitates a method that allows us to perform spectroscopy measurements on the object of interests.

In this regard, we developed a workflow powered by deep learning (DL) that enables automated study of IV behavior of specific objects in microscopy images. As shown in Figure 2b and c, this workflow includes a ML model that converts stream AFM image to segmented objects of interest. The coordinates of these objects can be extracted from this object image and transfer to AFM, in doing so, the microscope performs IV measurements at all coordinates. An example is shown in Figure 2c, a topography image can be converted to grain boundary image, where the grain boundary is the object of interest here; then, the coordinates of grain boundaries are shown as the blue spots and IV measurements will be performed at these blue spots.

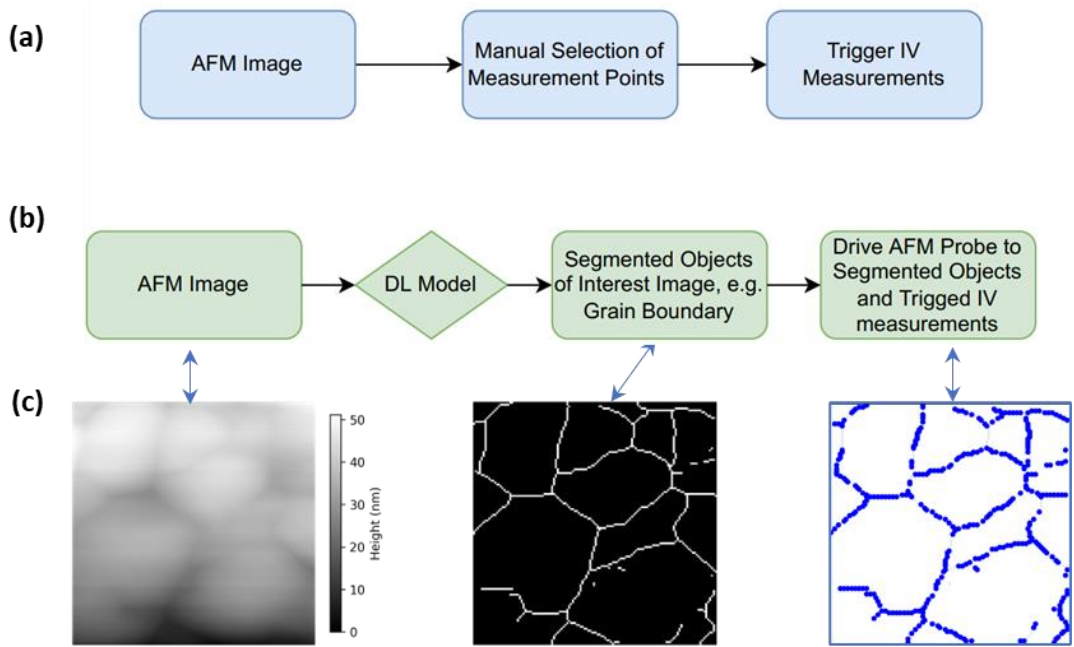

Figure 2. (a) standard spectroscopy measurement workflow, and (b) machine learning powered automated spectroscopy measurement (AE) workflow. The advantages of AE compared to standard method is, for instance, show in (c), AE can perform IV in all blue points shown in (c) but we can only measure a few points in standard method.

The DL model in the workflow is based on the modified version of holistically nested edge detector[46] augmented with residual connections. The original holistically nested edge detector contains five VGG-style[47] convolutional blocks, each with a side-output layer allowing learning features at multiple scales, which are fused at the end of the network. However, we found that for standard experimental SPM data, adding more than three convolutional blocks doesn't improve the domain boundaries detection rate and quality. On the other hand, replacing convolutional layers in each of the three convolutional blocks with a ResNet[47] module leads to a significant increase in the accuracy of domain boundaries (or walls) detection, as was demonstrated by some of the authors earlier.[48] We call this new neural network architecture ResHedNet, as shown in Figure 3a.

Finally, to improve the robustness of ResHedNet predictions on real-time data streams, we train an ensemble of ResHedNet models with different (pseudo-)random initialization of weights and different (pseudo-)random shuffling of training data batches. The data used for training the ResHedNet is shown in Figure 3b.

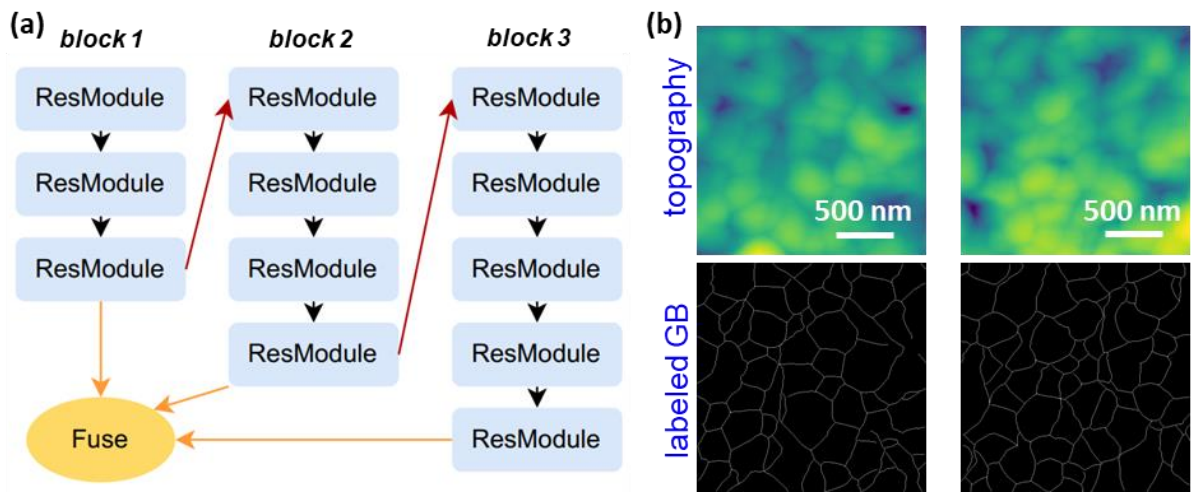

Figure 3. (a) ResHedNet network, where red arrows represent Max Pool, dark arrows represent Relu Activation, orange arrows represent transfer of tensors. The training in this work was done using focal loss. (b) the topography images and corresponding labeled GB images for training the ResHedNet network.

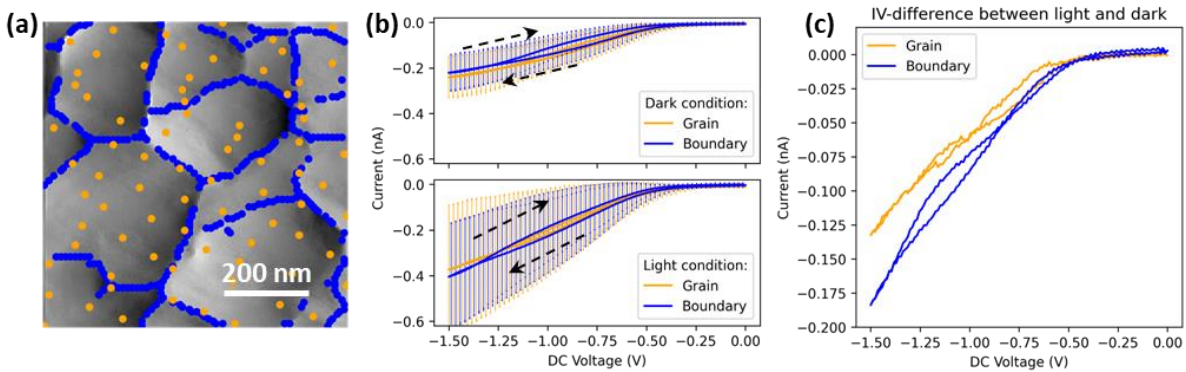

Figure 4. ResHedNet predicted grain boundary and performed IV at blue points. We modified the workflow to perform IV in grain as well, shown as orange points.

We trained the ResHedNet with pre-acquired topography image, the grain boundary (GB) in the topography *is* labeled with ImageJ for training. The training data is shown in Figure 3b. The training was performed using AtomAI software package and the detailed training process can be found in the provided Jupyter Notebook. Next, we implement the trained ResHedNet model in operating microscope to convert real-time topography image to GBs image. Then, the IV measurements were performed at GBs, and grain locations (which is the reverse of GB image). As shown in Figure 4a, the blue spots correspond to GBs and the yellow spots correspond to grains, where IV measurements were performed at each spot in an automated manner. Since MHP is a photovoltaic material, its property alters upon light illumination. Multiple works have indicated that light illumination change MHP property via reducing ion migration activation energy, photoinduced strain, phase segregation, etc. Therefore, we performed IV measurement at these locations under both dark condition and light illumination. Here the illuminating light is a built-in LED light in the microscope, its intensity is equivalent to 0.1 sun. Shown in Figure 4b is averaged IV response under dark condition at GB and grain. It is observed that the conductivity of GB under light is slightly smaller than grain, while the IV of GB exhibits a larger hysteresis. Under light condition, shown in Figure 4c the maximum current at -1.5 V indicates that GB is more conductive than grain, nonetheless, the IV hysteresis of GB is still larger.

To check the photo response of grain and GB, we also plotted the difference of IV under light and dark in Figure 4d, which are obtained by subtracting dark IV from light IV. The large error bar in Figure 4 implies that the GB behavior can be largely different and highlights the importance of investigating each GB point. Obviously, GB shows larger IV difference, suggesting that the GBs are more photo responsive. The discussion of GB effect on photovoltaic effect has been controversial, some believe the GB is beneficial for PV effect while others suggest GB is detrimental.[49, 50] Our results show that, on one hand, the GB exhibits larger photocurrent, on the other hand, GB exhibits larger IV hysteresis. The origin of these phenomena is likely due to ion migration and charge accumulation at GB.[51] The charge accumulation leads to high carrier density at GB under illumination and hence higher conductivity. Ion migration is due to the higher defect density at GB, and induces larger IV hysteresis under both dark and light. Accordingly, we suggest that the effect of GB cannot be simply evaluated by higher current, which may mislead us to conclude the beneficial effect of GB. Owing to the systematic investigation of IVs at GB here, we are able to observe the larger IV hysteresis at GB, indicating the detrimental side of GBs.

In addition to compare the average behavior of GBs and grains, we also explore the IV behavior as a function of location within GB or grain. To do so, we define several physical descriptors to represent the property of IV curve. As shown in Figure 5a, the turn on voltage represents the voltage where the current starts increasing; the hysteresis factor represents the difference between forward and reverse IVs; the maximum power represents the maximum power intensity under the sweeping voltage. These properties are extracted from each IV curve, and plotted as a function of locations. Shown in Figure 5b-c is the distribution of these properties at GBs, where the color represents the magnitude of corresponding physical descriptors. Interestingly, under dark condition, the turn on voltage around the GB junction points is extremely large (dark blue color in the first image in Figure 5b). There is also a trend that the turn on voltage gradually increases (from red color to cyan color) near the junction points. However, the behavior of turn on voltage is largely different under light as shown in the first image of Figure 5c—the GB junction points mostly do not show largest turn on voltage now, instead some GB middle (middle of GB) shows large turn on voltage. In the meantime, GB junction points also show smaller hysteresis and maximum power under dark, while these behaviors of GB junction points are also different under light condition.

In contrast to the clear trend of IV behavior at GB related to junction points, the IV behavior in grains is almost random, as shown in Figure 5d-e.

To understand the origin of the IV behavior at GB junction points, we check the exact IV curves at GB junction points. A mask is used to extract the locations with large turn on voltage under dark condition (according to our observation in Figure 5b, large turn on voltage locations are near GB junction points), the extracted locations are shown as red spots in Figure 6a. The corresponding averaged IV curves from these locations are plotted in Figure 6b. It is interesting that these locations are insulating under dark as evidenced by the flat IV curve (yellow curve in Figure 6b). However, these locations become highly conductive under light and exhibit large IV hysteresis (blue curve in Figure 6b).

Inspired by this observation, we further explored the surface potential of this sample by KPFM, in particular, the surface potential of GB junction points. The KPFM result is shown in Figure 6c, it is seen that GB exhibits higher surface potential than grains and the GB junction points exhibit further higher surface potential—higher than the GB middle. This is consistent with our observation in CL measurement, where we found the junction points exhibit higher intensity of component related to $PbI_2$. Since $PbI_2$ is known as a p-type semiconductor, it can lead to higher surface potential at the GB junction potential. However, several previous works indicated that excess $PbI_2$ in MHP leads to a morphology difference.[52-54] In our work, we did not observe distinct morphology difference, we suggest that this is most likely because of the small amount of $PbI_2$ in our film, which may buried in the MHP so that it does not change morphology, nevertheless it does affect the MHP properties and hence lead to the change in CPD, CL, and conductivity.

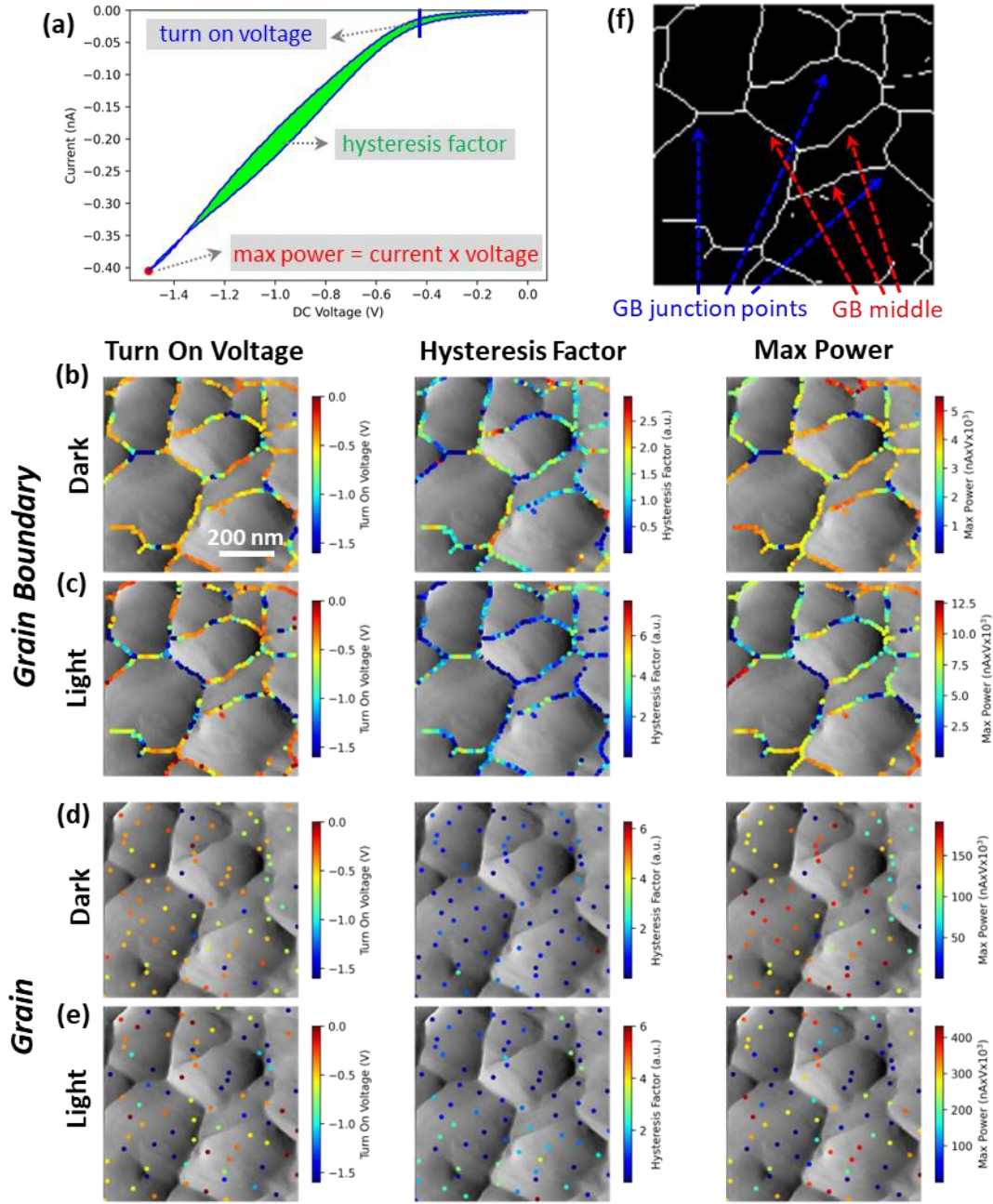

Figure 5. (a), three physical descriptors are defined, (b-e) shows the distribution of these descriptors as a function of location.

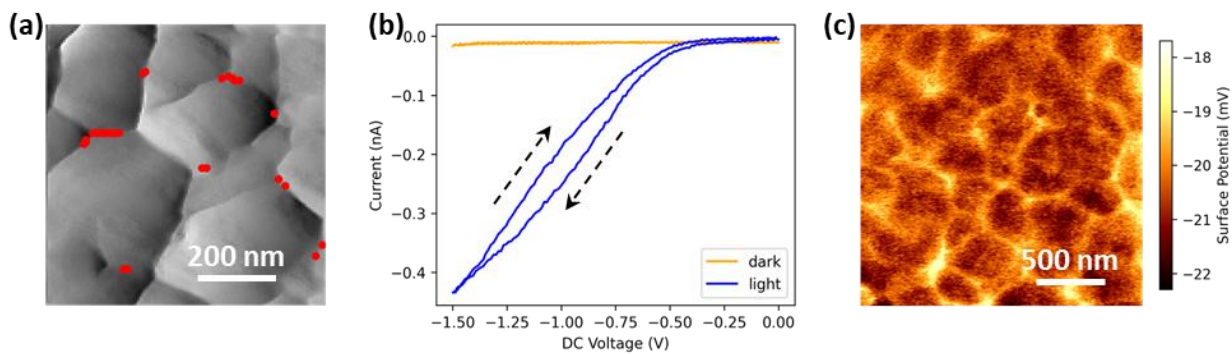

Figure 6. IV curves at junction points. Extract the insulating points under dark conditions, shown in a, and they indeed distributed at junction points. Then, we plot the averaged IV under dark and light conditions, show in b. It is seen that these points are insulating under dark, and show large hysteresis under light.

To our knowledge, even though GB and grain are extensively investigated, such behavior of GB junction points is discovered for the first time here. Tremendous works indicate that GB affects the degradation of MHP, so here a further question is how the GB junction points affect the stability of MHP. Photovoltaic devices operate under light and electric field, so it is worth testing the stability of GB junction points under these external field. Since such a microstructure cannot be tested on device level, we select microscope as our tool for stability study as well. Figure 7a shows the topography and GB coordinates for IV measurements. In this stability study, we apply multiple cycles of sweeping voltage waveform under light condition, as shown in Figure 7b. We used the evolution of IV curve as an indicator for stability, shown in Figure 7c is the evolution of averaged IV curves, where the maximum current at -1.5 V gradually increases as a function of time (from blue to red).

Same to previous analysis, we also extract the physical descriptors of turn on voltage, hysteresis factor, and maximum power from the IV curves. Here, in order to show the IV curve change between the beginning and the end, we plot the difference of physical descriptors between the 1$^{st}$ cycle and 10$^{th}$ cycle in Figure 8a-c. It is interesting that the turn on voltage at GB junction points almost does not change, indicated as cyan color (difference is 0 V) in Figure 8a. Correspondingly, the difference in hysteresis and maximum power is also near 0. We further created a mask to extract the points with small difference in turn on voltage, shown as the blue color spots in Figure 8d. It is seen that these spots mostly located near GB junction points. The averaged IV curve from these spots in Figure 8e is shown consecutively. First, we found that the current for these spots (near GB junction points) is smaller than other spots; second, the change of IV curves during continuous sweeping at these spots is also smaller than other spots.

This result indicates that the IV property of these GB junction points is relatively constant than other GB regions, implying that junction points do not induce significant degradation of MHP under light and in electric field during a short time scale (e.g., here the IV sweep time is only 12 s). This observation is consistent with our hypothesis that there is small amount of PbI$_2$ buried in MHP at GB junction points—several works have suggested that excess PbI$_2$ can passivate defects and hence suppress the ion migration,[55, 56] which results in relatively constant IV at GB junction points. However, the long-term effect of junction points and the composition of junction points on stability needs further investigation; and noteworthily, our CL results indicate blue-shifted peak near some GB junction points, such blue-shifted peak was attributed to a partially decomposed intermediate phase.[14, 57, 58] Nonetheless, in both cases, our study here suggested that the GB junction points can play critical roles in MHP stability.

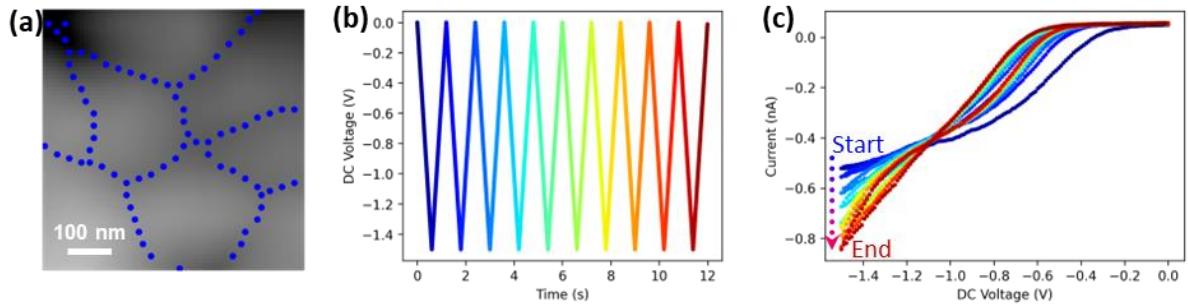

Figure 7. Evolution of IV under continuous sweeping voltages. (a) topography image and IV measurement points. (b) 10 cycles sweeping voltage waveform. (c) averaged IV at GBs.

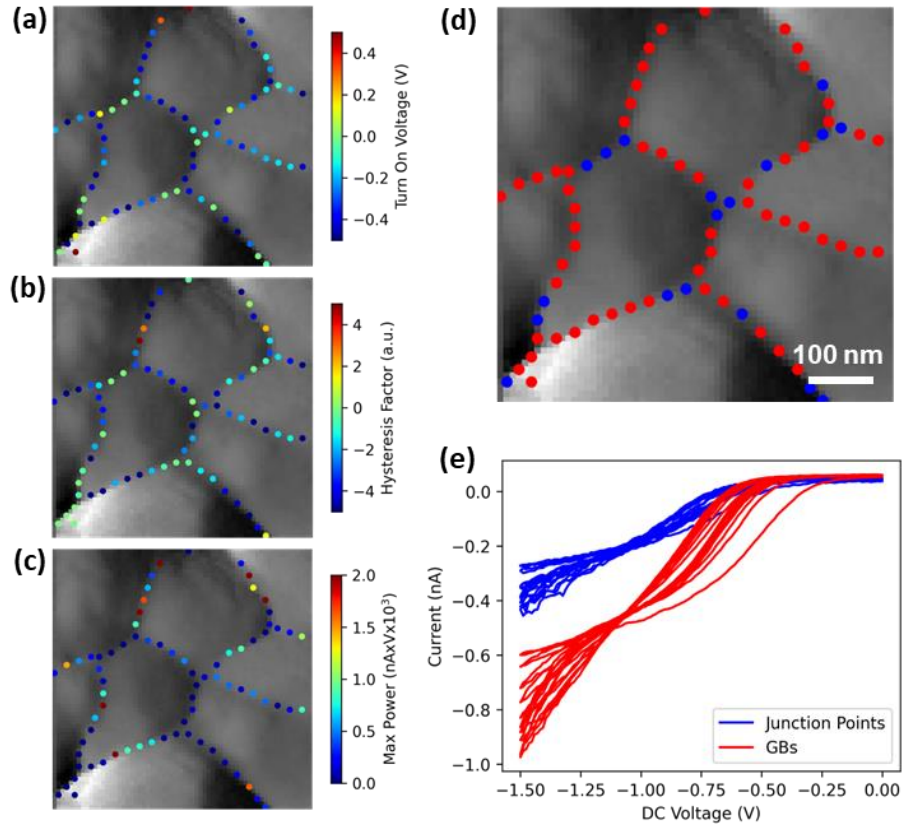

Figure 8. Evolution of IV characteristics under continuous sweeping voltages. (a)-(c) evolution of physical descriptors extracted from IV data shown in Figure 7, the difference between first sweep cycle and tenth sweep cycle of (a) turn on voltage, (b) hysteresis, (c) maximum power as a function of location, respectively. (d) markers of masked small change IV (blue points) and large change IV (red points). (e) averaged IV corresponding to the markers in (d).

In summary, we developed an automated conductive atomic force microscopy approach for investigating the local conductivity of functional materials. This approach combines the power of machine learning and AFM. We used this approach to study a MHP thin film. For the first time we revealed the behavior of GB junction points in MHP film, which are insulating under dark condition and exhibit larger current-voltage hysteresis under light conditions. Combined with CL measurement and KPFM measurement, we showed that the junction points have higher $PbI_2$ composition and hence higher surface potential. Our IV measurements suggest that the junction points are more photoactive and do not lead to significant degradation of MHP in a short timescale. This work highlights the role of GB junction points, while most previous works only focused on GB and grains. In addition, the developed automated microscopy can be universally used for investigations of other functional materials.

**Methods:**

*MHP synthesis:*

1.0M of $Cs_{0.17}FA_{0.83}PbI_3$ perovskite precursor solution was prepared by dissolving a mixture of respective volumetric amounts of $FAPbI_3$ and $CsPbI_3$ in a mixed solvent of DMF and DMSO [DMF (v) : DMSO (v) = 5:1] in a $N_2$ glovebox. The solution was spin-coated on the ITO substrates with a two-step process for making the film; 500rpm 10 seconds (1000 rpm/s) followed by 4000 rpm 35 seconds (2000 rpm/s), with the perovskite solution added before spin-coating. Chlorobenzene was gently added 10 s before the end of the second step. The films were then annealed at 150 °C for 10 min.

*Hyperspectral CL microscopy:*

A FEI Quattro SEM instrument with a Delmic Sparc CL collection module, utilizing a parabolic mirror to collect the CL signals was used. An electron beam with an acceleration voltage of 5 kV (with a beam current of 32 pA) was passed through a hole in the parabolic mirror for sample excitation, with an acquisition time of 400 ms per CL spectrum with a 60-nm size pixel; CL spectra were collected by scanning across the fixed area (defined with a size of 79×53 pixels). All measurements were conducted in a low vacuum environment (50 Pa $H_2O$ vapor) to mitigate sample charging. We note that such measurement conditions did not cause any observable sample degradation during measurement (over a week).

*cAFM and KPFM:*

Conductive AFM and KPFM measurements were performed in an Oxford Instrument Asylum Research Cypher microscope with Budget Sensor Multi75E-G Cr/Pt coated AFM probes (~3 N/m). The Cypher microscope is equipped with a National Instrument DAQ card and LabView script for applying voltage and acquiring IV data. The automated experiment workflow is designed in a Python Jupyter Notebook.

**Acknowledgements**

This effort (implementation in SPM, measurement, data analysis) was primarily supported by the center for 3D Ferroelectric Microelectronics (3DFeM), an Energy Frontier Research Center funded by the U.S. Department of Energy (DOE), Office of Science, Basic Energy Sciences under Award Number DE-SC0021118. This research (ensemble-ResHedNet and CL microscopy) was supported by the Center for Nanophase Materials Sciences (CNMS), which is a US Department of Energy, Office of Science User Facility at Oak Ridge National Laboratory. J.Y. and M.A. acknowledge support from National Science Foundation (NSF), Award Number No. 2043205. The authors acknowledge Rama K. Vasudevan for necessary developments for AE-SPM.

**Conflict of Interest Statement**

The authors declare no conflict of interest.

## Authors Contribution

Y.L. and S.V.K. conceived the project. M.Z. developed ensemble-ResHedNet. Y.L. implemented ensemble-ResHedNet in cAFM and obtained results. J.Y. and A.M. synthesized MHP film and conducted CL measurement. All authors contributed to discussions and the final manuscript.

## Data Availability Statement

The method that supports the findings of this study are available at https://github.com/yongtaoliu/MHP_GB_by_ResHedNet.